\documentclass[12pt, fleqn]{article}
\usepackage{amsmath,amssymb,amsfonts}
\usepackage{graphics}
\usepackage[pdftex]{hyperref}
\usepackage{array}

\date{\small February 15, 2005}
\title{Inequality and Growth: A Two-Player
Dynamic Game with Production and Appropriation}
\author{ Julio Huato\footnote{St. Francis College, Brooklyn, New York.  E-mail: jhuato@sfc.edu.}}

\begin{document}
\maketitle

\newpage

\tableofcontents

\newpage
\linespread{1}
\begin{abstract}
\noindent This paper models a two-agent economy with production and
appropriation as a noncooperative dynamic game, and determines its
closed-form Markovian Nash equilibrium. The analysis highlights the
parametric conditions that tip the economy from a nonaggressive or
``cooperative'' equilibrium to outright distributional conflict.
The model includes parameters that capture the role of appropriation
technology and destructiveness. The full dynamic implications of the
game are yet to be explored, but the model offers a promising
general framework for thinking about different technological and
economic conditions as more or less conducive to cooperation or
distributional conflict.

\end{abstract}

\newpage

\newpage
\linespread{1}

\section{Introduction}

Empirical studies conducted on various data sets show that, after
controlling for other relevant factors, larger initial economic
disparities in an economy are associated with lower subsequent
economic performance.\footnote{A summary of the ongoing empirical
debate on this matter is provided in section (\ref{litrev}).}

Motivated by the finding of these empirical studies, this paper lays
out a two-player differential game model of a simple economy with
production and appropriation, determines its Markovian Nash
equilibrium analytically, and examines its implications.  Key
insights about the dynamics of this artificial economy are gleaned
from these results.

Although the nonlinearity of the Markovian Nash equilibrium's
canonical system of differential equations may preclude a
closed-form solution of its dynamic allocation problem, numerical
solutions can be implemented. However, such numerical solutions and
their analysis are outside the scope of this paper.

By \emph{production}, it is meant that each player can use the stock
of wealth under her possession to create a flow of net new wealth
through a simple concave technology. \emph{Appropriation} is defined
as the set of costly actions undertaken to protect one's own
possessions and/or to grab someone else's possessions.

In this context, ownership means degree of possession or
disposability of wealth.  In the practice of social life, legal and
ethical rules, or rather the degree to which these are enforced,
appear to determine ownership.  But ownership in the sense used in
this paper does not require legal or ethical recognition. It is
entirely the outcome of the initial distribution of possessions,
technological conditions of production and appropriation, and
economic outcomes (shadow prices and actions chosen optimally over
time) that flow from constrained utility maximization.\footnote{The
model includes as parameters the fraction of each player's wealth
that is subject to appropriation, the ``productivity'' of
appropriation outlays, and the fraction of the wealth taken away
from the victim that winds up effectively appropriated by the
aggressor. Each of these parameters can be viewed as reflecting
institutional (political, legal, or ethical) conditions in the
economy.  However, this model does not require that. They can also
be viewed as entirely determined by technology and economics.} Thus,
ownership is viewed in a continuum, with different degrees of force.
In this light, the extent of appropriation can be interpreted as a
negative or inverse measure of wealth \emph{excludability}, as the
force of ownership is called in the theoretical
literature.\footnote{The model in this paper is consistent with both
``political economy'' and ``social conflict'' causality mechanisms
as per the classification in Ferreira (1999). More generally,
although the distinctions matter in practice, the concern here is
not whether appropriation is conducted legitimately, lawfully (e.g.,
fiscal policy, eminent domain expropriation), or otherwise (e.g.,
blatant robbery).}

This model highlights the inefficiencies resulting from
appropriation and, in that sense, describes a mechanism by which a
skewed initial distribution of wealth may lead to distributional
conflict, large welfare losses, and thereby poor economic
performance.\footnote{Other mechanisms can explain the negative
relationship between initial wealth distribution and poor economic
performance in the framework of a dynamic game. One such mechanism
could be a players' difference in time preferences specified as
different rates of subjective discounting in the payoff function.}
But, this is hardly a surprising result. More importantly, the model
provides a way to examine the effects of different economic settings
-- each with its corresponding system of incentives -- on concurrent
growth, where these economic settings can be proxied by boundary
conditions and the relations among technological parameters.

In a one-period model, Grossman and Kim (1995) postulate
appropriation in the form of two distinctive actions: defense (which
they call ``fortifications'') and attack (which they call
``predation''). Their specification of the appropriation function,
including parameters that capture the efficacy and destructiveness
(wastefulness) involved in appropriation actions, has some
attractive features. Appropriation actions enter the appropriation
function inversely (hyperbolically).  That is, the amount of her own
wealth a player manages to retain is an increasing and smooth
function of the appropriation actions she undertakes.  On the other
hand, the amount of player 2's wealth that player 1 manages to
appropriate is a decreasing and smooth function of player 1's
appropriation actions. Furthermore, the marginal products of
appropriation (in the form of wealth retained and grabbed) are
decreasing. Although slightly simplified, Grossman and Kim's basic
set-up is adopted in the model presented in this paper.\footnote{The
distinction between defensive and aggressive actions in Grossman and
Kim (1995) is abandoned to ensure a closed-form Markovian Nash
equilibrium.} In this sense, this model provides a dynamic
generalization of these authors' static results.

From a welfare analysis perspective, the natural Pareto-optimal
baseline is that of a cooperative dynamic game, where the players
agree to cooperate and share the dynamic efficiencies thus gained
prior to their playing the game.\footnote{Also known in the
literature as the ``coordination game.'' Section \ref{litrev}
summarizes this well-known case.} In such case, the game effectively
becomes a regular optimal control problem of the type studied in
Ramsey's (1928) and Koopmans's (1965) models. However, the
institutional setting that enables the players to make prior binding
cooperative agreements is necessarily exogenous. At first sight,
under the assumption of self-interested players, the cooperative
case appears to have little practical relevance, since the
institutional setting that induces cooperation is not fully
explicated within the model.

The advantage of parameterizing the efficacy and destructiveness of
appropriation is precisely that it allows us to explicate how such
conditions may be brought about endogenously by two self-interested
players in strategic interaction (e.g. individuals, classes,
nations) with no \emph{a priori} motivation to cooperate. In other
words, we seek the conditions under which -- given an initial
distribution of wealth -- self-interested players may optimally
renounce appropriation actions, thus reaping and sharing the dynamic
gains from cooperation, in the face of constraints in the technology
and economics of production and appropriation. Grossman and Kim
(1995) call this type of solution to the game a \emph{nonaggressive
equilibrium}.\footnote{In Grossman and Kim's model, the
nonaggressive equilibrium case is clearly distinguished from the
coordination or cooperative case, because the former requires
positive defensive outlays to the point of deterring all offensive
expenditures, whereas the latter needs no appropriation outlays
whatsoever. In other words, the latter is Pareto optimal whereas the
former is Pareto suboptimal. However, in the model in this paper, a
nonaggressive equilibrium entails zero appropriation outlays,
effectively exhibiting the same Pareto superior properties of the
cooperative case.  For this reason only, at considerable semantic
cost, in this paper, the terms \emph{nonaggression} and
\emph{cooperation} are used interchangeably.}

The theorem establishing the existence of Markovian Nash equilibria
in differential games states that concavity in the controls of the
Hamiltonian, which in turn requires concavity of the felicity
function and of the state equation of motion, is sufficient to
ensure the existence of at least one Markovian Nash equilibrium of
the right sign, e.g. a ``supremum'' of the objective
functional.\footnote{See Dockner et al. (2000) for the statement and
proof of the existence theorem.} But, determining Markovian Nash
equilibria in practice is a vexing task. In particular, nonlinear
interactions between state and control variables makes it extremely
difficult to obtain a closed-form Markovian Nash
equilibrium.\footnote{In a two-player game, the Markovian Nash
equilibrium is obtained from the first-order conditions for each
instantaneous (``static'') optimum for both players, by solving
these conditions simultaneously for both players' strategies in
terms of the state and costate variables and parameters. A Nash
equilibrium is a strategy that maximizes the payoff of each player
under the assumption that the other player simultaneously chooses
her optimal strategy. For each player, a Nash equilibrium is the
best response to the Nash equilibrium strategies of the other
player. Markovian strategies (also called ``closed-loop'' or
``feedback strategies'') are dynamic decision rules in which the
choice of a player's current action depends entirely on the current
state vector and the current time. The Nash equilibrium in a game
played with Markovian strategies is said to be a Markovian Nash
equilibrium.} This is undoubtedly the biggest analytical roadblock
in solving a differential game model and deriving clear economic
implications from it. This is perhaps the chief reason why this
method has so far attracted limited interest among economic
theorists.

The unique contribution of this paper is the specification of a
model that, without sacrificing the appealing nonlinearity in the
interactions between actions and wealth in the appropriation
function (and, thereby, in the equation of motion and the
Hamiltonian) that Grossman and Kim (1995) suggested, has a
\emph{unique} closed-form Markovian Nash equilibrium that -- in
spite of its apparent complexity -- yields significant insight into
the economics of appropriation and growth.

Section \ref{frame} sets up the model in its more general form in
the framework of a differential game. Section \ref{analysis}
analyzes the equilibrium conditions. Section \ref{litrev} discusses
relevant literature. Section \ref{concl} has a few concluding
remarks.

\section{Game} \label{frame}

This section is organized as follows: Subsection \ref{frame} sets up
the model. Subsection \ref{noncoop} finds the point-in-time
conditions for an individual player's optimal path, given feasible
strategies chosen by the other player. Subsection \ref{markov}
determines the general Markovian Nash equilibrium of the
noncooperative game.

\subsection{Framework} \label{frame}

Consider a single-good economy with two utility-maximizing players
$i = 1, 2$ over a period from the present $t=0$ to a terminal time
$t=T$. For each player $i$, the payoff functional is the sum of the
instantaneous utility (\emph{felicity}) of current consumption over
the (continuous time) life of the game:
\begin{eqnarray}
W_i = \int_0^T u[c_i(t)] \ dt \label{objfn} \\
\end{eqnarray}

\noindent where $u'(c) > 0$ and $u''(c) < 0$.  That is, felicity
$u[c(t)]$ is assumed to be a continuous, smooth, increasing, and
strictly concave function of current (instantaneous) consumption,
which is assumed to be nonnegative. For simplicity, there is no
discounting of future utility.\footnote{Ramsey (1928), Pigou (1932),
and Harrod (1948) view the discounting of utility of future
generations as ethically indefensible.  Subjective discounting does
not alter the overall results in this model, since the discount
factor can be factored out of the Hamiltonian after redefining the
costate in terms of (loosely speaking) current ``utils,'' rather
than present-value-at-zero ``utils.'' The only difference lies in
the time interpretation of the costate.}

Each player $i=1, 2$ is initially endowed with a given amount of
wealth:
\begin{eqnarray}
x_i(0) = x_{i0} \label{initcond}
\end{eqnarray}

The initial endowments of wealth need not be equal for both players.
In fact, as it will be established below, the difference in the
initial endowments of wealth between the players, is a key parameter
in the model.

Using existing wealth as the sole input, at each point in time, each
player can produce new wealth according to the following production
function:
\begin{eqnarray}
y_i(t)= y[x_i(t)]
\end{eqnarray}

\noindent where $ y'(x) > 0$ and $y''(x) < 0$.  In words, the
production function is assumed to be a continuous, smooth,
increasing, and strictly concave function of the stock of existing
wealth.  The stock of existing wealth is used as the input of the
production function at each $t \in [0, T]$.  It is assumed that all
outlays of wealth for current consumption and other current actions
are deductions from the flow of wealth produced at that point in
time.

Besides using output to consume and/or produce for the next instant,
each player may engage in \emph{appropriation} actions, namely
instantaneous outlays of wealth made to protect her existing wealth
and/or grab the other player's existing wealth.  It is also assumed
that, at each point in time $t \in [0,T]$, under existing
technology, only a fixed fraction $0 < \delta < 1$ of each player's
existing stock of wealth is effectively subject to appropriation. In
other words, the current (instantaneous) flows of production,
consumption, or appropriation outlays are not subject to
appropriation.

Appropriation actions are denoted by $a_i(t)$ and assumed to be
nonnegative.\footnote{From this point on, the time reference in
parenthesis is omitted to streamline the notation.}

At $t \in [0, T]$, the fraction of her own wealth player $i$ retains
(the \emph{retention rate}) is:
\begin{eqnarray}
p_i = \frac{1}{1 + \theta a_j}
\end{eqnarray}

\noindent for $i \neq j = 1, 2$, where $\theta > 1$ is a constant
parameter that reflects the efficacy of appropriation actions.
Clearly, $0 \leq p_i \leq 1$.  More specifically, $\lim_{a_j
\rightarrow 0} p_i=1$ and $\lim_{a_j \rightarrow \infty} p_i=0$.
Also, $p_i$ is a monotonically decreasing function of $a_j$: $d
p_i/d a_j =  - \theta/(1 + \theta a_j)^2< 0$.

The portion of the other player's wealth a given player can
effectively grab is:
\begin{eqnarray}
\gamma (1 - p_j) = \gamma \Big(\frac{ \theta a_i}{1 + \theta
a_i}\Big)
\end{eqnarray}

\noindent for $i \neq \ j = 1, 2$, where $0 < \gamma <1$ is a
constant parameter (the \emph{salvage ratio}) that captures the
portion of wealth lost to appropriation by one of the players that
is actually appropriated by the aggressor. In other words, only a
fraction $\gamma$ of the portion of wealth that player $j$ loses is
effectively appropriated by player $i$.  Somehow the fraction
$(1-\gamma)$ winds up as a deadweight loss in the process. It
follows from the result about $p_i$ that $(1 - p_j)$ is a
monotonically increasing function of $a_i$.

The instantaneous rate of change of player $i$'s stock of wealth is
given by:
\begin{eqnarray}
\dot{x}_i = y(x_i) - c_i - a_i - \delta x_i + \delta x_i \Big(
\frac{1}{1+ \theta a_j} \Big) + \gamma \delta x_j \Big( \frac{\theta
a_i}{1 + \theta a_i} \Big) \label{eqnmot}
\end{eqnarray}

\noindent where $\dot{x} \equiv dx/dt$.  This is called the
\emph{equation of motion} of player $i$'s wealth.  In words, the
rate of change of a player's wealth equals the output of current
production, current consumption, and appropriation outlays, minus
the portion of wealth subject to appropriation, plus the portion of
one's own wealth retained and the portion of the other player's
wealth appropriated.

The equation of motion can be expressed in terms of the retention
rates and simplified further:
\begin{eqnarray}
\dot{x}_i = y(x_i) - c_i - a_i - \delta x_i ( 1 - p_j) + \gamma
\delta x_j (1 - p_i) \label{stateeqnmot}
\end{eqnarray}

This being a one-good economy, the exchange of the good at each $t
\in [0, T]$ would be pointless.  However, the value of each unit of
wealth transferred to the next instant may mean different amount of
``utils'' to each player.  In other words, the instantaneous shadow
price of each player's wealth may differ. With intertemporal
exchange allowed between the players (i.e. borrowing and lending),
the two players could in principle trade the good intertemporally up
to the point where the instantaneous shadow prices of wealth are
equalized.

To simplify matters and train the focus strictly on appropriation,
this model does not allow intertemporal exchange between the
players. To frame this assumption differently, any discrepancy
between the instantaneous shadow price of player $i$'s wealth and
player $j$'s wealth can only translate into appropriation actions
between them.

Each player's dynamic problem is to choose a strategy (i.e. a set of
time paths for her controls over the life of the game $\{c^*_i(t),
a^*_i(t)\}_0^T$ that maximizes her payoff functional, subject to her
own wealth equation of motion, boundary conditions, nonnegativity
constraints, and the strategy chosen by the other
player.\footnote{The states and controls are assumed to be
nonnegative.} Player $i$'s individually-optimal strategy is a
function of both players' existing stock of wealth.\footnote{In
other words, this is a perfect-information differential game.
Strategies that depend only on the current state of the system are
known as \emph{closed-loop} controls or Markovian strategies.}

The transversality condition is given by:
\begin{eqnarray}
x_i(T) \lambda_i(T) = 0 \label{transv}
\end{eqnarray}

\noindent for at least one of the players $i = 1, 2$, where $T$ is
given exogenously or is to be determined endogenously. In words, the
transversality condition says that the game ends when either the
wealth of (at least one of) the players or its instantaneous shadow
price is zero.  The case of a given terminal time requires that the
terminal instantaneous shadow price of wealth be pushed down to
zero, since wealth has no use after $T$.  If one player loses all
her wealth, the game turns into a regular Ramsey-type optimal
control problem for the surviving player.

\subsection{The first-order conditions} \label{noncoop}

It is assumed that the players have no prior binding agreement to
cooperate, i.e. each of them acts independently of the other to
maximize her individual payoff functional (\ref{objfn}).

Player $i$'s full Hamiltonian function for each $t \in [0, T]$ is:
\begin{eqnarray}
H_i = u(c_i) + \lambda_{ii} \Big[y(x_i) - c_i - a_i - \delta x_i
\Big( \frac{\theta a_j}{1+ \theta a_j} \Big) + \gamma \delta x_j
\Big(
\frac{\theta a_i}{1 + \theta a_i} \Big)\Big] \nonumber \\
+ \lambda_{ij} \Big[y(x_j) - c_j - a_j - \delta x_j \Big(
\frac{\theta a_i}{1+ \theta a_i} \Big) + \gamma \delta x_i \Big(
\frac{\theta a_j}{1 + \theta a_j} \Big)\Big] \label{ham}
\end{eqnarray}

Unlike a regular optimal control problem, in the differential-game
framework, two costates can be specified for player $i$.  One of
them, $\lambda_{ii}$ captures the marginal effect on the objective
functional of relaxing $i$'s own equation of motion by one unit of
wealth.  It is the instantaneous shadow price or gross return rate
of a unit of the player's own wealth transferred to the next
instant, expressed in utility units of the current instant.  More
simply, $\lambda_{ii}$ is called the \emph{own} instantaneous shadow
price. The second costate variable, $\lambda_{ij}$, captures the
marginal effect on $i$'s objective functional that results from a
unit change in player $j$'s wealth dynamic constraint. It is called
here the \emph{cross} instantaneous shadow price.

The first-order necessary conditions for an optimal consumption and
appropriation strategy for player $i$ at each $t \in [0, T]$,
holding constant the strategy chosen by player
$j$,\footnote{Technically, the strategy chosen by player $j$ is
restricted to a set of feasible strategies.  In this model, the
feasibility of a player's strategy is given by its nonnegativity,
given the nonnegativity of the state variables.} are the same
conditions required to maximize $i$'s Hamiltonian (\ref{ham}):
\begin{eqnarray}
u'(c_i) = \lambda_i \label{foc1}\\
\lambda_i - \lambda_i \gamma \delta x_j \Big[ \frac{\theta}{1 +
\theta a_i} - \frac{\theta^2 a_i}{(1 + \theta a_i)^2} \Big] =
\lambda_j \delta x_j \Big[ \frac{\theta}{1 + \theta a_i} -
\frac{\theta^2 a_i}{(1 + \theta a_i)^2} \Big] \label{foc2}
\end{eqnarray}

\noindent at $t \in [0, T]$ and for $i \neq j = 1, 2$. The detailed
derivation of these conditions is shown in the Appendix.

Note that, since $x_i$ affects $H_i$ and $H_j$ in the same way,
through the equations of motion of $x_i$ and $x_j$, then $- \partial
H_i/\partial x_i = - \partial H_j/\partial x_i = \dot{\lambda}_{ii}
= \dot{\lambda}_{ji} = \dot{\lambda}_{i}$. Also, $-
\partial H_j/\partial x_j = - \partial H_i/\partial x_j =
\dot{\lambda}_{jj} = \dot{\lambda}_{ij} = \dot{\lambda}_{j}$.
Therefore, $\lambda_{ii} = \lambda_{ji} = \lambda_{i}$ and
$\lambda_{jj} = \lambda_{ij} = \lambda_{j}$ by construction of the
Hamiltonians and the transversality condition.

The second-order sufficient conditions for a maximal value of the
Hamiltonian and, thereby, of player $i$'s payoff functional at each
point in time are satisfied by the strict concavity of the felicity
function and of the equation of motion in both $c_i$ and $a_i$,
since -- assuming the instantaneous price of wealth is nonnegative
-- a linear combination of two concave functions is itself a concave
function.

\subsection{The Markovian Nash equilibrium} \label{markov}

A \emph{Markovian Nash equilibrium} reconciles the first-order
conditions for both players consistent with their respective
first-order necessary conditions, at each $t \in [0, T]$. A
Markovian Nash equilibrium is defined as the optimal consumption and
appropriation strategy of a player under the assumption that the
other player also chooses her own optimal strategy. If a Markovian
Nash equilibrium exists and the players are at such an equilibrium,
then they have no incentive to move away from it for a given set of
parameters and boundary conditions.

The Markovian Nash equilibrium can be found by simultaneously
solving the system of equations formed by the first-order conditions
of both players, for the control variables (current consumption $c$
and appropriation outlays $a$), in terms of the parameters $\delta,
\theta, \gamma$, the stocks of wealth of the players $x_i, x_j$, and
the respective instantaneous shadow prices $\lambda_i, \lambda_j$.

By Pontryagin's maximum principle in the context of a differential
game, the solution to the game's dynamic allocation -- i.e. the
optimal time paths for the stocks of wealth of both players and for
their respective instantaneous shadow prices -- must satisfy the
Markovian Nash equilibrium at each $t \in [0, T]$.

The general Markovian Nash equilibria of the game is given by:
\begin{eqnarray}
c^*_i = u'^{-1}(\lambda_{i}) \label{cstar} \\
a^*_i = \frac{\sqrt{\theta \delta x_j (\gamma \lambda_{i} -
\lambda_{j})/\lambda_{i}} - 1}{\theta} \label{aistar}
\end{eqnarray}

\noindent for $i \neq j = 1, 2$ and for each $t \in [0, T]$, where
$u'^{-1}(.)$ indicates the implicit inverse of the marginal utility
function (assuming it exists), the $\pm$ sign preceding the
square-root operator is eliminated by using the nonnegativity
conditions imposed by assumption on the control and state variables,
and the asterisk denotes point-in-time optimality.  The detailed
derivation of the Markovian Nash equilibria equations is shown in
the Appendix.

\section{Analysis} \label{analysis}

This section analyzes the Markovian Nash results derived in the
previous section. Subsection \ref{optc} interprets the condition for
the optimal consumption path. Subsection \ref{distrconf} sorts out
the conditions for a nonaggressive versus an aggressive equilibrium.
Some of the qualitative dynamics of the game are examined in
subsection \ref{dynamics}. Subsection \ref{coop} refers the
well-known single-agent optimal-control case.

\subsection{Optimal consumption} \label{optc}

To examine the implications of the first-order conditions and the
Markovian Nash equilibrium (which satisfies them for both players
simultaneously), consider the \emph{elasticity of substitution}
between consumption at two discrete points in time, $t$ and $s$:
\[ \sigma[c_t] \equiv - \frac{u'(c_s)/u'(c_t)}{c_s/c_t} \frac{d(c_s/c_t)}{d\{u'(c_s)/u'(c_t)\}}\]

The well-known result that the limit of this elasticity as the
difference between $s$ and $t$ becomes arbitrarily small yields
\[\sigma[c(t)] = - \frac{1}{c(t)} \ \Big(\frac{u'[c(t)]}{u''[c(t)]}\Big)\]

\noindent is shown in the Appendix.  This expression says that the
point or \emph{instantaneous} elasticity of substitution, which
turns out to be the reciprocal of the negative of the
\emph{elasticity of the marginal utility}, $\epsilon(c) \equiv
[u''(c) c]/u'(c)$, a measure of the curvature of the felicity
function.  Thus, the following expressions for the growth rate of
consumption $\hat{c} \equiv \dot{c}/c$ can be derived substituting
into condition (\ref{foc1}):
\begin{eqnarray}
\hat{c^*}_i = - \frac{1}{\epsilon} \hat{\lambda_{i}} = \sigma(c^*_i)
\hat{\lambda_{i}} \label{foc1b}
\end{eqnarray}

Note that $\sigma(c_i) \geq 0$, because felicity is increasing in
$c_i$ and strictly concave, i.e. $\epsilon \leq 0$.  Hence, in this
form, condition (\ref{foc1}) says that the relation between the
growth rate of consumption and the growth rate of the instantaneous
shadow price of wealth is nonnegative in a proportion determined by
the instantaneous elasticity of substitution between consumption now
and consumption in the next instant. This is a well-known result in
the modern theory of economic growth.

In terms of the elasticity of marginal utility, condition
(\ref{foc1}) says that if felicity is more rather than less curved
(i.e. if growing consumption yields a rapidly declining marginal
utility) and consumption is strictly positive, then a player's
growth rate of consumption will lag behind the growth rate of the
instantaneous shadow price of her own wealth.

\subsection{Distributional conflict} \label{distrconf}

On the other hand, using the retention rate $p$ defined above,
condition (\ref{foc2}) simplifies nicely to:
\begin{eqnarray}
\frac{a^*_i}{(p^*_i)^2 - p^*_i} = \delta x_j \Big(\frac{\lambda_{j}-
\gamma \lambda_{i}}{\lambda_{i}}\Big)
\end{eqnarray}

To derive the parametric conditions under which $a_i \geq 0$,
L'H\^opital's rule can be used to take the limit of this expression
as $a_i$ approaches zero from the right:
\begin{eqnarray}
\lim_{a^*_i \rightarrow 0^+} \Big( \frac{a^*_i}{(p^*_i)^2 -
p^*_i}\Big) = -\frac{1}{\theta}
\end{eqnarray}

After some algebra, we obtain the following condition for $a_i \geq
0$:
\begin{eqnarray}
\frac{\gamma \lambda_{i} - \lambda_{j}}{ \lambda_{i}} \geq
\frac{1}{\theta \delta x_j} \label{ai0}
\end{eqnarray}

The detailed derivation of these results appears in the Appendix.

In words, player $i$'s optimal appropriation actions at a given
point in time increase above zero as, holding all other constant,
(1) the difference between player $i$'s instantaneous shadow price
of own wealth and instantaneous shadow price of cross wealth, (2)
the salvage ratio, (3) the efficacy of appropriation, or (4) the
fraction of $j$'s wealth subject to appropriation increase above a
certain point.

Equilibrium condition (\ref{aistar}) or, equivalently, inequality
(\ref{ai0}) suggest that, at a given $t \in [0, T]$, for player $i$
to move from zero to positive appropriation actions, she requires
\emph{sizeable} incentives in the form of relatively large
differences between $\gamma \lambda_{i}$ and $\lambda_{j}$. Larger
values of $\delta$ or $\theta$ can only scale up difference ($\gamma
\lambda_{i}- \lambda_{j}$), but it is this difference itself that
determines the algebraic sign of the argument in the square root
operator of equilibrium condition (\ref{aistar}). And this argument
is required to be larger than one for $a_i > 0$.  Less than large
differences between the instantaneous shadow prices are insufficient
to induce player $i$ to employ any portion of her wealth in
appropriation actions.

In the context of this model, parameters $\delta$, $\theta$, and
$\gamma$ can be viewed as reflecting exogenous technological
conditions.  On the other hand, given the salvage ratio, the ratio
$(\gamma \lambda_{i} - \lambda_{j})/\lambda_{i}$ is driven
endogenously by the dynamics of the instantaneous shadow prices.
This raises the question of what characterizes such dynamics.

\subsection{The game's dynamics} \label{dynamics}

To examine qualitatively the dynamics of the game, consider the
Euler equation for player $i$:\footnote{The Euler equation of each
player results from equalizing the time derivative of the
instantaneous shadow price of player $i$'s wealth and the negative
of the partial derivative of the Hamiltonian with respect to the
player's stock of wealth.}
\begin{eqnarray}
\dot{\lambda}_{i} = \lambda_{i} \Big[ \delta \Big( \frac{\theta
a^*_j}{1 + \theta a^*_j} \Big) - y'(x_i) \Big] - \lambda_{j} \delta
\gamma \Big( \frac{\theta a^*_j}{1 + \theta a^*_j} \Big)
\label{eulereq}
\end{eqnarray}

In terms of the retention rate, the growth rate of the instantaneous
shadow price simplifies to:
\begin{eqnarray}
\hat{\lambda_{i}} = \delta (1 - p^*_j) - (\lambda_{j}/\lambda_{i}) \
\delta \gamma (1 - p^*_j) - y'(x_i) \label{grlambda}
\end{eqnarray}

The Euler equation is known in the economic-growth literature as the
Keynes-Ramsey rule.  In the context of this game, equation
(\ref{eulereq}) or (\ref{grlambda}) could be called the
\emph{modified} Keynes-Ramsey rule and be interpreted as saying that
the marginal rate of substitution between consumption now and
consumption at the next instant must equal the marginal rate of
transformation of wealth now into wealth at the next instant via
production and/or appropriation.  In the case of production, the
marginal output is valued at the player's own instantaneous shadow
price.  In the case of appropriation, the retained output of
appropriation is valued at the player's own instantaneous shadow
price, but the grabbed output of appropriation is valued at the
cross instantaneous shadow price.

Substituting equation (\ref{grlambda}) into condition (\ref{foc1b})
above, the optimal growth rate of player $i$'s consumption can be
expressed in terms of $\delta$, $\theta$, $\gamma$, $p_j$, her own
marginal product $y'(x_i)$, and the ratio of cross to own
instantaneous shadow prices of wealth:
\begin{eqnarray}
\hat{c}^*_i = \sigma(c^*_i) \Big\{ \delta ( 1 - p^*_j) \Big[\frac{
\lambda_{i} - \gamma \lambda_{j} }{\lambda_{i}} \Big] - y'(x_i)
\Big\} \label{grcons}
\end{eqnarray}

Clearly, if player $i$ chooses to reduce one small portion of her
wealth $dc_i$ now and transfer it via production or appropriation in
order to have more wealth for consumption at the next instant, the
``utils'' lost now will be $dc_i \ u'(c_i)$.  On the other hand, the
``utils'' gained will be the sum of gains in ``own utils' $dc_i \{[1
+ (y'(x_i) + \delta p_j]/[1 + \delta] \}$ plus ``cross utils'' $dc_i
[\delta \gamma (1 - p_j)]$.\footnote{Remember that player $i$'s
instantaneous valuation of her own wealth coincides with player
$j$'s instantaneous valuation of $i$'s wealth. So ``cross utils''
does not mean ``utils'' in the subjective sense of the other player
only, but also own ``utils'' \emph{as if} the one player had the
same level of consumption as the other.} Along the optimal path, the
``utils'' lost and gained must be equal or else the path is not
optimal.

Equation (\ref{grcons}) also implies that the difference between the
instantaneous shadow prices (as a fraction of own's instantaneous
shadow price), a crucial factor in tipping the players' optimal
strategies from zero to positive appropriation actions, is directly
related to the dynamics of consumption and inversely related to the
marginal product and the size of appropriation:\footnote{At any
stationary or steady state consistent with the Markovian Nash
equilibrium conditions (i.e. the game's optimal dynamic equilibrium
path), the ratio $(\lambda_{i} - \gamma \lambda_{j})/\lambda_{i}$
also turns out to be a measure of the inequality in the distribution
of wealth at each point in time.  See the discussion below.}
\begin{eqnarray}
\frac{ \lambda_{i} - \gamma \lambda_{j} }{\lambda_{i}} =
\frac{\hat{c^*}_i/\sigma(c^*_i) - y'(x_i)}{\delta ( 1 - p^*_j)}
\end{eqnarray}

If $a_i \rightarrow 0$, equation (20) and (21) turn into:
\begin{eqnarray}
\hat{\lambda}_i|_{a_i \rightarrow 0} = - y'(x_i) \\
\hat{c}^*_i|_{a_i \rightarrow 0} = - \sigma(c^*_i) \ y'(x_i)
\label{grcons2}
\end{eqnarray}

\noindent and equations (\ref{grcons2}), one per player,
characterize entirely the dynamics of the optimal nonaggressive time
path of the game's economy.  In other words, the outcome of the game
is the same as the outcome of the coordination or cooperative game.
\footnote{See section \ref{coop}.}

The canonical system of differential equations whose solutions are
required to determine the optimal time paths of wealth for each
player, as well as the optimal time paths of the shadow prices, is
given by the equation of motion (\ref{eqnmot}), Euler equation
(\ref{eulereq}), given initial conditions $x_i(0) = x_{i0}$, and
transversality condition (\ref{transv}) for both players $i=1,2$.
Valued for the Markovian Nash equilibrium, the latter becomes:
\begin{eqnarray}
x_i(T) u'[c^*_i(T)] = 0
\end{eqnarray}

This condition says that, if the instantaneous value of the marginal
utility at the terminal point $T$ is positive, then it is not
optimal for a player to keep a positive stock of wealth at that
point, since she could increase the value of her objective
functional by consuming it.

From the analysis above, it is clear that the stationary states in
the game, defined by $\dot{c}^*_i = \dot{a}^*_i = \dot{x}_i=0$, must
satisfy the following conditions:
\begin{eqnarray}
 y'(x_i)= \delta ( 1 - p^*_j) \Big[\frac{\lambda_{i} - \gamma \lambda_{j}
  }{\lambda_{i}} \Big] \label{eureka0} \\
\hat{\lambda}_i = \hat{\lambda}_j - \hat{x}_j \Big(\frac{\gamma
\lambda_i - \lambda_j}{\lambda_j} \Big) \label{eureka} \\
y(x_i) + \gamma \delta (1 - p^*_i) = c^*_i + a^*_i + \delta x_i (1 -
p^*_j) \label{eureka2}
\end{eqnarray}

\noindent for each player $i = 1, 2$ over the life of the game $[0,
T]$.

The first condition follows from equation (\ref{grcons}), which in
turn results from substituting the Euler equation (\ref{eulereq})
into first-order condition (\ref{foc1}).  The second condition for
stationarity results from taking the time derivative of Markovian
Nash solution (\ref{aistar}) and setting it equal to zero.  The
third condition follows from the wealth equation of motion
(\ref{eqnmot}).

It is trivially true that a stationary state of the game is given by
$c^*_i = a^*_i = x^*_i = 0$ for at least one of the players. More
importantly, since by assumption $y'(x_i) < 0$ and $y''(x_i) > 0$,
equation (\ref{eureka0}) implies that, at any steady state, player
$i$'s stock of wealth and her wealth's instantaneous shadow price
are negatively related. Moreover, equation (\ref{eureka}) implies
that the growth rate of player $i$'s own instantaneous shadow price
is positively related to the player $j$'s stock of
wealth.\footnote{The detailed proof of these two assertions is shown
in the Appendix.}

In words, at a steady state satisfying the Markovian Nash
equilibrium conditions (i.e. a dynamic optimal equilibrium), the
higher the stock of wealth of a player, the lower her wealth's
instantaneous shadow price, and the higher her level of current
consumption.  And, vice versa, the lower the stock of wealth of a
player, the higher her wealth's instantaneous shadow price, and the
lower her level of current consumption.\footnote{This is the game's
analog of Ramsey's model condition (\ref{lambdadot}) below.}

This should not be a surprising result.  In this game, aside from
their difference in the (given) initial endowments of wealth, the
players are otherwise identical. The values of the parameters apply
for both players. Felicity, the production function, and the
appropriation function are all the same for both players, smooth and
strictly concave. Also, each player's optimal appropriation actions
depend only on the other player's stock of wealth at $t$.

Therefore, in the dynamic solution of the game, the ratio $(\gamma
\lambda_{i} - \lambda_{j})/\lambda_{i}$, the decisive factor that
tips the economy from its nonaggressive equilibrium to outright
distributional conflict, reflects not only a difference between the
players in their respective instantaneous shadow prices, but also
the inequality in their levels of current consumption and in their
stocks of wealth. In other words, this ratio is \emph{a measure of
the wealth inequality }prevailing in the economy.

The explicit determination of interior dynamic solutions for this
game requires further work.  Given the nonlinearity of the canonical
system of differential equations of the game, computing specific
numerical solutions, i.e. based on specific forms of the felicity
and production functions and boundary conditions, may be required to
examine in more detail the comparative dynamics of the game.

\subsection{The coordination game} \label{coop}

This section summarizes results that apply to the game if, before
the game, the players strike a binding cooperative agreement whereby
they eliminate appropriation outlays and share the game by
coordinating their respective consumption rates.  The following will
consider this case.  However, these same results apply to (1) the
degenerate case of the differential game when the boundary
conditions and relations between the parameters do not ensure the
strict inequality case in (\ref{ai0}), i.e. when $a_i =0$, and to
(2) any of the players individually, if the other player's stock of
wealth or its own shadow price approach zero.

The economy's state and control variables are:
\begin{eqnarray}
x = \pi x_1 + (1-\pi) x_2 \\
c = \pi c_1 + (1-\pi) c_2 \\
a = \pi a_1 + (1-\pi) a_2
\end{eqnarray}

\noindent where $\pi$ is the agreed-upon weight of player $i$ in the
economy, assumed constant across the welfare function, the
distribution of wealth, and the division of actions.

Assuming a simple additive welfare function, both players coordinate
the choice of their control trajectories to maximize the joint
objective functional:
\begin{eqnarray}
W = \int_0^T u[c(t)] d t \label{cooppayoff}
\end{eqnarray}

\noindent for $t \in [0, T]$, subject to the boundary conditions,
the nonnegativity constraints, and the following equation of motion:
\begin{eqnarray}
\dot{x} = y(x) - c \label{coopstateeqnmot}
\end{eqnarray}

For the players, maximizing equation (\ref{cooppayoff}) is
equivalent to maximizing the Hamiltonian:
\begin{eqnarray}
H =  u(c) + \lambda [y(x) - c]
\end{eqnarray}

\noindent at each $t \in [0, T]$, since $\delta$ and $a_i$ for $i=1,
2$ are zero by agreement.

This is a streamlined version of Ramsey's (1928) model of growth,
which has been studied extensively.\footnote{The extra simplifying
assumptions adopted in this paper (zero depreciation and zero
population growth) lead to a golden rule level of production that is
at once the optimal and the maximal rates of consumption.} For
reference purposes, the first-order condition for a maximum value of
the functional at each point in time is:
\begin{eqnarray}
u'(c) = \lambda
\end{eqnarray}

The time path of the aggregated stock of wealth in the economy is
given by the equation of motion and the time path of the
instantaneous shadow price of wealth is given by:
\begin{eqnarray}
\dot{\lambda} = - \lambda y'(x) \label{lambdadot}
\end{eqnarray}

By Pontryagin's maximum principle, the optimal paths of consumption
and wealth $\{ c^*(t), x^*(t)\}_0^T$ must satisfy the system of
differential equations formed by (\ref{coopstateeqnmot}) and
(\ref{lambdadot}) valued at the first-order condition.  Given a
boundary condition on initial wealth $x(0) = x_0$ and on the
terminal instantaneous shadow price of wealth $\lambda(T) = 0$, the
system can be solved for the optimal paths of wealth and its shadow
price.

Perhaps a more intuitive description of the dynamics of the system
can be attained by taking the time derivative of the costate in the
point-in-time first-order condition, substituting in the state and
costate differential equations, and using the instantaneous
elasticity of substitution $\sigma(c) \equiv - u'/(u'' c)$, so that
the system of differential equations becomes:
\begin{eqnarray}
\dot{c}= \sigma(c) \ c \ y'(x) \\
\dot{x} = y(x) - c
\end{eqnarray}

Thus, the dynamics of consumption and wealth accumulation in this
economy can be summarized by contrast to the steady or stationary
state where $\hat{c}=\hat{x}=0$ (in which the levels of consumption
$\bar{c}$ and wealth stock $\bar{x}$ are constant over time) and
summarized as follows:
\begin{eqnarray}
\hat{c} > 0 \ \textrm{if} \ y'(x) > 0 \ \textrm{i.e. if} \ x < \bar{x} \\
\hat{c} < 0 \ \textrm{if} \ y'(x) < 0 \ \textrm{i.e. if} \ x > \bar{x} \\
\hat{c} = 0 \ \textrm{if} \ y'(x) = 0 \ \textrm{i.e. if} \ x =
\bar{x}
\end{eqnarray}
\begin{eqnarray}
\hat{x} > 0 \ \textrm{if} \ y(x) > c \\
\hat{x} < 0 \ \textrm{if} \ y(x) < c \\
\hat{x} = 0 \ \textrm{if} \ y(x) = c
\end{eqnarray}

\section{Literature} \label{litrev}

This section discusses narrow slivers of literature relevant to the
model in this paper.  Subsection \ref{thref} refers to some
predecessors of the model in this paper. Subsection \ref{empdeb}
summarizes the empirical research that motivated the model.
Subsection \ref{randomth} closes the section with references and
thoughts on the notion of cooperation between self-interested
economic agents.

\subsection{Theoretical references} \label{thref}

The key theoretical predecessor for this paper is Grossman and Kim's
(1995) pioneering two-player static game.  The model in this paper
owes to Grossman and Kim the specification of the appropriation
function as a hyperbolic function.  However, this paper modifies the
set-up of the appropriation function in a crucial respect: it
abandons the distinction between defensive and offensive actions.

The decision to alter Grossman and Kim's specification of the
appropriation function can be justified on two grounds.  First, not
much is lost by shedding the distinction.  Arguably, the key element
in the specification is the smooth concavity of the appropriation
function.  The second reason is tractability. With defense specified
as a type of action separate from attack, the first-order necessary
condition for an individual optimum that determines the optimal
level of appropriation actions (defense and attack) become dependent
on the player's own wealth, aside from depending on the other
player's wealth.  This turns the derivation of a Markovian Nash
equilibrium into a combinatorial nightmare.  On the other hand,
without the distinction, the Markovian Nash conditions are easy to
pin down.

Doubtlessly, the determination of the Markovian Nash equilibrium is
the biggest analytical roadblock in solving differential game models
and deriving clear economic insight from them. This is perhaps the
main reason why this method has so far attracted little interest
among economic theorists.  This is remarkable, since early
theoretical results and applications of differential game theory
date back to the mid 1950s and early 1960s. Isaacs (1965) compiles
the seminal papers. Friedman (1971) provides a systematic and
general description of the mathematics of differential games,
including the theorem of the existence of a Markovian Nash
equilibrium (at the time referred as ``the saddle point solution'').
Intriligator (1971) introduces differential games to students of
economics. Dockner et al. (2000) has a modern version of the
existence theorem.

Simpler (more tractable) forms of specifying the appropriation
function seem much less interesting or have been tried by other
authors. Lancaster (1973) and Hoel (1978) use the differential game
approach to highlight what Lancaster calls the ``dynamic
inefficiency of capitalism.''  Their specification allows them to
obtain elegant and straightforward ``bang-bang'' solutions to their
models. They achieve simplicity by ruling out extra-economic
appropriation -- and the nonlinearities that arise from its
specification. They postulate ``workers'' and ``capitalists'' as
players engaged in regular exchange in the labor and product
markets.  Since their definition of ``capitalism'' is such that
capitalists own all the physical wealth and the workers only their
labor power, Lancaster's and Hoel's papers are similar to mine
insofar as they all show that a dynamic welfare loss ultimately
results from inequality. The difference is that, in their models,
the inequality between workers and capitalists translates not into
appropriation, but into a separation between the current consumption
(and thereby saving) and investment decisions \emph{\`{a} la}
Keynes.

To judge by the topical literature, engineering and operations
research specialists are rather comfortable with numerical
approximate solutions to \emph{both} the Markovian Nash equilibria
and the solution to the dynamic allocation problem.\footnote{For a
rare paper using numerical approximation to compute both the
Markovian Nash equilibrium and the dynamic solution to an economic
differential game, see Itaya (2000).}  The theoretical economic
literature is far from having adopted this attitude. Instead, it
usually expects sharp analytical solutions and qualitative and --
even -- quantitative predictions that can be demonstrably reconciled
with the tested tenets of basic economic intuition. While there is
some acceptance of numerical solutions to the dynamic allocation
problem proper, the suspicion remains.  The concern seems to be that
-- given the complexity of the relations involved -- the numerical
approximation of a Markovian Nash equilibrium, when fed into the
derivation of the dynamic allocation solution, could limit or even
distort the outcome to the point of rendering it useless.

\subsection{The empirical debate on inequality and growth} \label{empdeb}

Galor and Zeira (1993) are credited with reviving interest in the
causal mechanism between initial inequality and subsequent growth.
The empirical finding that larger initial income inequality is
associated with lower subsequent growth is due to Person and
Tabellini (1994) and Alesina and Rodrik (1994). B\'{e}nabou (1996)
and Perotti (1996) surveyed the empirical literature and reported
that most studies found the same result.

In a 1997 paper (published in the \emph{American Economic Review}),
Forbes (2000) challenged this finding and, using a new data panel
developed by Deininger and Squire (1996), reported instead a
positive relation between initial income inequality and subsequent
growth. Li and Zou (1998) also claimed to have found a positive
relation. Deininger and Squire (1998) themselves conducted an
analysis of their data set and found that the negative relation was
robust. Sz\'ekely and Hilgert (1999) questioned the quality of the
Deininger and Squire data for Latin America and showed that Forbes'
results might have been dependent upon the method used to compute
inequality. Birdsall and Londono (1997) found a negative relation
between initial human-capital inequality and subsequent growth.

Barro (1999), analyzing an extended data set, found ``little overall
relation between income inequality and rates of growth and
investment.'' However, ``higher inequality tends to retard growth in
poor countries and encourage growth in richer places.''  Since most
countries in the world are classified as poor and most people in the
world live in poor countries, it is apparent that the former
conclusion has a larger relevance in global welfare than the latter.
Lucas (1988) has suggested compellingly that there are massive gains
in efficiency to be made in the world economy by helping poor and
large economies to grow.\footnote{``Is there some action a
government of India could take that would lead the Indian economy to
grow like Indonesia's or Egypt's? If so, \emph{what}, exactly? If
not, what is it about the `nature of India' that makes it so?  The
consequences for human welfare involved in questions like these are
simply staggering: Once one starts to think about them, it is hard
to think about anything else.'' Lucas (1988).}

Sokoloff and Engerman (2000) and Engerman and Sokoloff (2002)
marshalled a large body of historiographic information to draw a
contrast between Latin America and Anglo America, where the former
started off with a more skewed distribution of its factor
endowments, which in turn led it to evolve weaker legal and
political institutions and, through them, experience a poorer
economic performance. Easterly (2001a) and Easterly (2001b) used a
careful econometric specification to validate Engerman and
Sokoloff's hypothesis.  While the challenge by Forbes (2000) and Li
and Zou (1998) was based on panel regressions, Panizza (2002)
analyzed a cross-state data panel for the United States.  He found
no evidence of a positive relation and ``some evidence'' of a
negative relation, although sensitive to the method used to measure
inequality.\footnote{The initial inequality considered in the
empirical literature refers to consumption, income, and ownership of
an array of assets, physical, human, and even intangibles.  Some
studies include markers such as race, ethnicity, language, gender,
or age, where such markers lead to social ranking. The conceptual
and empirical relations among physical wealth, human wealth, income,
and consumption are reasonably well established. In this
paper, we refer to `wealth inequality' in the broadest sense.} \\

\subsection{Random thoughts on cooperation} \label{randomth}

It is common in the general equilibrium literature to contrast the
decentralized markets result to that derived without appeal to
separation theorems, i.e. by invoking the existence of some sort of
``social planner.'' In the welfare analysis of general equilibrium,
the social planner results are used as the baseline to compare the
results obtained for the decentralized markets case. Implicit in the
exercise is the notion that an economy in which all self-interested
economic agents perfectly coordinate their individual actions is
superior to any conceivable alternative.\footnote{For an example of
this practice, see Stokey and Lucas (1989), chapter 1.} More
realistic economic arrangements (e.g. decentralized markets) can
only aspire to match the welfare heights that such a perfect economy
would attain if it could only come about.

In spite of the highly abstract -- in fact, heroic -- character of
the assumptions underpinning general equilibrium (perfect
information, perfect markets, convexity of preferences and
technology, and zero externalities), the decentralized markets case
appears as much more realistic, more robust, and more likely to
emerge out of actual historical economic evolution than perfect
coordination as personified in the abstract role of the ``social
planner.'' After all, the chief role of the social planner, i.e. to
provide a ``consistent'' (not self-contradictory) mapping from
individual utility to the welfare function, is left completely
unexplained.

Arrow (1963) looked carefully into this mapping from individual
utility functions to a general welfare function and derived very
stringent conditions for such a mapping to exist and for the
resulting welfare function to meet some basic ``rationality''
criteria. One of the conditions is, precisely, the existence of a
``dictator'' able to form a consistent welfare function, thus
getting around the ``impossibility'' of interpersonal utility
comparisons.  The ``social planner'' can thus be viewed as a sort of
``benevolent dictator.''

Arrow's Impossibility Theorem notwithstanding, welfare and economic
policy analysis continue to depend on the notion of social welfare.
Considering how pervasive redistributive actions (e.g., fiscal
policy) are, in virtually every economy, it is clear that the
criterion of Pareto efficiency, by eluding distributive judgments,
is too restrictive to be of practical use in welfare economics and
policy analysis. In most cases, policymakers face re-distributive
tradeoffs associated with different economic policy choices. It is
rarely the case that there is a clearly Pareto-superior
choice.\footnote{Stiglitz (2003) claims that facing gray-area
situations with no evident Pareto-improving policies in sight is the
rule in actual policymaking.} It is no surprise then that, in the
absence of uncontroversial results (similar to Pareto efficiency) to
guide welfare economists and policymakers in redistributive
dilemmas, theorists are forced to forge more restrictive welfare
notions and even flirt with the old cardinalism.\footnote{On the
former, see Bardhan, Bowles, and Gintis (1998), who replace the
notion of ``Pareto efficient,'' unable to judge the welfare effects
of nonmarket-mediated asset redistribution, with that of
``productivity enhancing,'' which denotes welfare gains from
nonmarket-mediated asset redistribution, in situations where
informational imperfections lead to incomplete or
too-costly-to-enforce contracts. On the latter, see for instance
Layard (2003).}

Other strands of the literature suggest how the idea of cooperation
in the strategic interaction between economic agents may be an
optimal response. For instance, one way to view the cooperative game
is as the outcome of a merger of the two players' interests in order
to internalize the external costs from their nonmarket-mediated
interaction (i.e., appropriation). In our model, intertemporal
exchange between the players is excluded by assumption. Therefore, a
formal real interest rate does not emerge. However, this does not
mean that prices are absent. The instantaneous \emph{shadow prices}
implicit in production and appropriation are in the nature of what
Coase calls ``prices in their widest sense.'' Thus, the negative
externalities of appropriation can be viewed as large-scale Coasian
transaction costs that might be better handled by a merger of the
players' interests into a single economic entity.  Examples of this
are ``firms'':
\begin{quote}
It is clear that an alternative form of economic organization which
could achieve the same result at less cost than would be incurred by
[bargaining between private owners] would enable the value of
production to be raised. [\ldots] [T]he firm represents such an
alternative to organizing production through [\ldots] transactions
[between private parties]. Within the firm, individual bargains
between the various co-operating factors of production are
eliminated and for a [\ldots] transaction [between private owners]
is substituted an administrative decision. [\ldots] In effect,
[\ldots] the firm would acquire the legal rights of all the parties,
and the rearrangement of activities would not follow on a
rearrangement of rights by contract but as a result of an
administrative decision as to how the rights should be used.'' Coase
(1960, pp. 115-116).
\end{quote}

However, in this case as well, the precise mechanism by which the
merger is designed and implemented remains exogenous.

Sen (1997) argues elegantly that a Pareto superior solution to the
prisoner's dilemma game would be achieved by self-interested players
if they behaved paradoxically, i.e., by turning it into an
``assurance game'' (under a principle of reciprocity, players give
each other prior assurances that they will not shirk) or under the
assumption of ``socially conscious'' play (each player prefers to do
the right thing whether or not the other does the same):
\begin{quote}
That the Prisoner's Dilemma could disappear if people had different
preferences is true but hardly interesting. What is, however, quite
significant is the fact that even if the people involved continued
to have the same Prisoner's Dilemma type preferences, but behaved as
if their preferences were as in the Assurance Game (or better still
\emph{as if} they had [\ldots] ``socially conscious'' preferences
[\ldots]), they could be better off \emph{even in terms of their
true preferences}. This is precisely where the question of cultural
orientation comes in, and it may provide a social case for
encouraging values that reorient a person's choices and actions even
if his personal welfare functions remain unaltered. In a sense, this
is a matter of morality, and there are of course many other spheres
of life as well in which a society throws up moral values that
attempt to dissociate choice from individualistic rational calculus.
Indeed, this is a common phenomenon for ``homely virtues'' like
honesty, keeping promises, etc., but what is important to recognize
here is the relevance of all this to the problem of work motivation
and therefore to income distribution.\footnote{Sen (1997), pp.
98-99.} \label{senftnt}
\end{quote}

Sen's scenario is that of prisoner's dilemma games in which
self-interested players remain self-interested in their felicity
function while behaving \emph{as if} it (their felicity function)
were altruistic. This is a more sophisticated view than that of
invoking the altruistic behavior of the players as a \emph{deux ex
machina}.\footnote{There is a growing literature on the evolution of
altruistic preferences and cooperation from first economic
principles, partly spanned by Axelrod's (1984) work.  For an
example, see Bowles and Gintis (2000).}    But, still, it requires,
rather implausibly, that the set of preferences individuals use to
subjectively map their actions to their wellbeing be different by
assumption from the preferences that guide their actual behavior in
the face of constraints.

Sen's reflections are a reply to Marx, whose description of ``pure''
communism is that of a ``coordinated-game'' type of society in which
common ownership excludes formal exchange. Marx held the view that
the emergence of a communist society was necessary because the
``socialization of production'' (and life in general) would
increasingly require the ``socialization of ownership.''  By the
phrase ``socialization of production,'' he meant an increasing
interdependence between producers, mainly due to technological
progress.\footnote{``Let us imagine an association of free men,
working with the means of production held in common, and expending
their many different forms of labor-power in full self-awareness as
one single social labor force.'' (1976, p. 171.) ``Within the
collective society based on common ownership of the means of
production, the producers do not exchange their products; just as
little does the labor employed on the products appear here as the
value of these products, as a material quality possessed by them,
since now, in contrast to capitalist society, individual labor no
longer exists in an indirect fashion but directly as a component
part of total labor.'' (1938, p. 85.) ``On the basis of communal
production, the determination of time remains, of course, essential.
The less time the society requires to produce wheat, cattle, etc.,
the more time it wins for other production, material or mental. Just
as in the case of an individual, the multiplicity of its
development, its enjoyment and its activity depends on the
economization of time. Economy of time, to this all economy
ultimately reduces itself. Society likewise has to distribute its
time in a purposeful way, in order to achieve a production adequate
to its overall needs; just as the individual has to distribute his
time correctly in order to satisfy the various demands on his
activity. Thus, economy of time, along with the planned distribution
of labor time among the various branches of production, remains the
first economic law on the basis of communal production. It becomes
law, there, to an even higher degree. (1973, p. 172-173.)}  Marx
seemed to have envisioned the emergence of a pure communist society,
without formal exchange, as the result of a long transition that
required the suppression of the main forms of economic inequality
and the gradual development of a new ethics of cooperation and
publicly-minded outlook.\footnote{A brief discussion of the
transitional process leading to pure communism is in Marx (1938).
Sen (1995) provides an erudite and meticulous discussion of the
difficulties of the notion of economic and/or social equality.}

In modern terms, Marx envisioned technological progress as a process
leading to the emergence of ever larger and more pervasive
externalities that markets, based on private ownership, would be
unable to handle efficiently.  The agency for the transition to
communism would be the ``propertyless direct producers,'' the
workers, who would have the least vested interest in the
\emph{status quo}.\footnote{Another way to frame Marx's view of
technological progress in modern terms is as a process by which an
ever larger set of goods becomes \emph{public}. A public good is
both \emph{nonrivalrous} (on the extreme, \emph{nondepletable}) and
\emph{nonexcludable}. While the nonrivalrous (and nondepletable)
character of goods depends on their physical or technical
attributes, their excludability is socially conditioned. As
suggested above, excludability is another name for the ability of
individuals or their agencies to enforce their ownership rights and,
in the terms of our model, is inversely related to the extent of
appropriation.} However, the specifics of this transitional process
are missing in Marx's writings or vaguely described as a process of
workers' collective self-growth through their struggle against the
capitalists. It is not clear the extent to which Marx understood the
major sources of moral hazard arising from the common disposition of
productive wealth, let alone how to deal with them in
practice.\footnote{Private ownership can be viewed as a rough form
of private insurance in the face of economic uncertainty. Similarly,
public ownership can be viewed as social insurance aimed to
eliminate unsystematic risk. For a modern treatment of the topic of
social insurance, see Shiller (1996).}

Keeping in mind that the notion of cooperation used in this paper
does not distinguish explicitly between direct cooperation and
cooperation mediated by exchange in markets,\footnote{In fact, free
voluntary exchange through markets (as opposed to appropriation)
could be viewed as an organized form of cooperation. On the other
hand, direct voluntary cooperation, in which reciprocity is not
regulated by prices or even enforced, can be viewed as an informal
mode of exchange.} a loose interpretation of the implications of the
model above is that -- at least when considering a given point in
time -- the conditions that enable self-interested players to
cooperate may be much more robust that usually thought.  Again, at
least for a given point in time, distributional conflict appears to
be the exception.

\section{Conclusions} \label{concl}

This paper presented a dynamic model of a two-agent economy with
production and appropriation using the framework of differential
games. The closed-form Markovian Nash equilibrium of this economy
was determined and information about the economics of appropriation
and growth was extracted from it.

Given initial conditions and parameter values, for a given point in
time, the model implies that appropriation is unlikely, requiring as
it does \emph{extreme} conditions of inequality in wealth
possession, levels of consumption, and -- therefore -- instantaneous
shadow prices of wealth.

However, the full dynamic implications are not yet clear.  One
possibility is that, as a result of the strict concavity of the
appropriation function, the system has a self-correcting mechanism
built in that prevents initial disparities from exploding to the
point where one of the players losses all her wealth.

Another possibility is that the nonlinearities in the canonical
system of differential equations, that the dynamic allocation
optimal paths must satisfy, build up initial small disparities in
wealth possession (and shadow prices) over time to the point of
creating conditions that induce distributional conflict.  And
clearly, the introduction of randomness in the model would make it
more likely to generate distributional conflict situations.

The full solution of the dynamical allocation problem of the game,
and the analysis of its implications, is left for future work.

\newpage
\section*{Appendix} \addcontentsline{toc}{chapter}{Appendix
\dotfill}

\textbf{Retention ratio}

I define the retention ratio as:
\begin{eqnarray} p_i \equiv 1/(1 + \theta a_i)\end{eqnarray}

Then, the derivative of the retention rate with respect to
appropriation is:
\begin{eqnarray} \frac{dp_i}{da_i} = - \frac{\theta}{(1 + \theta a_i)^2}\end{eqnarray}
\begin{eqnarray} \frac{dp_i}{da_i} = - \Big(\frac{\theta }{1 + \theta a_i}\Big)
\Big(\frac{1}{1 + \theta a_i}\Big) \Big(\frac{a_i}{a_i}
\Big)\end{eqnarray}
\begin{eqnarray} \frac{dp_i}{da_i} = - \frac{(1-p_i) p_i}{a_i} \end{eqnarray}
\begin{eqnarray} \frac{dp_i}{da_i} = \frac{p^2_i-p_i}{a_i} \label{d_pi}\end{eqnarray}

On the other hand, the derivative of the loss rate with respect to
appropriation is:
\begin{eqnarray} \frac{d(1-p_i)}{da_i} = \frac{(1 + \theta a_i) \theta - \theta a_i \theta }{(1 + \theta a_i)^2} \label{d1_pi0}\end{eqnarray}
\begin{eqnarray}\frac{d(1-p_i)}{da_i} =  \frac{\theta + \theta^2 a_i - \theta^2
a_i}{1 + \theta a_i} \end{eqnarray}
\begin{eqnarray} \frac{d(1-p_i)}{da_i} = \frac{\theta }{(1 + \theta a_i)^2} \end{eqnarray}
\begin{eqnarray}\frac{d(1-p_i)}{da_i} =  \frac{(1-p_i) p_i}{a_i} \end{eqnarray}
\begin{eqnarray}\frac{d(1-p_i)}{da_i} =  \frac{p_i-p^2_i}{a_i}
\label{d1_pi}\end{eqnarray}

\textbf{First-order conditions}

The Hamiltonian is defined as:
\begin{eqnarray} H_i = u(c_i) + \lambda_{ii} [y(x_i) - c_i - a_i - \delta x_i
( 1- p_j) + \gamma \delta x_j (1-p_i) \Big)] \nonumber \\ +
\lambda_{ij} [y(x_j) - c_j - a_j - \delta x_j (1 - p_i) + \gamma
\delta x_i (1 - p_j)]
\end{eqnarray}

The first-order necessary conditions for the Hamiltonian to be
maximized for each player and $t$ are derived as follows.  First,
with respect to consumption:
\begin{eqnarray} \frac{\partial H_i}{\partial c_i} = u'(c_i) - \lambda_{ii} =0\end{eqnarray}

Therefore:
\begin{eqnarray} u'(c_i) = \lambda_{ii} \end{eqnarray}

And then with respect to appropriation:
\begin{eqnarray} \frac{\partial H_i}{\partial a_i} = \lambda_{ii} [-1 + \gamma \delta x_j \frac{\partial (1-p_i)}{\partial a_i}] + \lambda_{ij} [-\delta x_j \frac{\partial (1-p_i)}{\partial a_i }] = 0\end{eqnarray}

Using equation (\ref{d1_pi}) above, one can derive a nice-looking
expression for this first-order condition:
\begin{eqnarray} \lambda_{ii} \Big[\gamma \delta x_j \Big(\frac{p_i-p^2_i}{a_i}\Big)\Big]  - \lambda_{ii} - \lambda_{ij} \Big[\delta x_j \Big(\frac{p_i-p^2_i}{a_i}\Big)\Big] = 0\end{eqnarray}
\begin{eqnarray} \lambda_{ii} \Big[\gamma \delta x_j \Big(\frac{p_i-p^2_i}{a_i}\Big)\Big]  - \lambda_{ij} \Big[\delta x_j \Big(\frac{p_i-p^2_i}{a_i}\Big)\Big] = \lambda_{ii} \end{eqnarray}
\begin{eqnarray} \delta x_j \ (\gamma \lambda_{ii} - \lambda_{ij} ) \Big(\frac{p_i-p^2_i}{a_i}\Big) = \lambda_{ii} \end{eqnarray}
\begin{eqnarray} \frac{a_i}{p_i-p^2_i}= \delta x_j \frac{ (\gamma \lambda_{ii} - \lambda_{ij} ) }{\lambda_{ii}} \label{foc2} \end{eqnarray}

\textbf{Substitution and marginal utility elasticity}

The instantaneous elasticity of substitution is the reciprocal of
the negative of the elasticity of the marginal utility.

To see this, consider two points in time $t$ and $s$.  The
elasticity of substitution of consumption between them is defined
as:
\begin{eqnarray} \sigma(c_t) \equiv - \frac{u'(c_s)/u'(c_t)}{c_s/c_t} \frac{d(c_s/c_t)}{d[u'(c_s)/u'(c_t)]} \end{eqnarray}

Take the limit of the expression above as $s \rightarrow t$:
\begin{eqnarray} \lim_{s \rightarrow t} \sigma(c_t) = \sigma[c(t)] = - \lim_{s \rightarrow t} \frac{u'(c_s)/u'(c_t)}{c_s/c_t} \times \lim_{s \rightarrow t} \frac{d(c_s/c_t)}{d[u'(c_s)/u'(c_t)]} \end{eqnarray}
\begin{eqnarray} \sigma[c(t)] = - \frac{ \lim_{s \rightarrow t} [u'(c_s)/u'(c_t)]}{\lim_{s \rightarrow t} [c_s/c_t]} \times \lim_{s \rightarrow t} \frac{d(c_s/c_t)}{d[u'(c_s)/u'(c_t)]} \end{eqnarray}
\begin{eqnarray} \sigma[c(t)] = - \frac{1}{1} \times \lim_{s \rightarrow t} \frac{d(c_s/c_t)}{d[u'(c_s)/u'(c_t)]} \end{eqnarray}
\begin{eqnarray} \sigma[c(t)] = - \frac{\lim_{s \rightarrow t} d(c_s/c_t)}{\lim_{s \rightarrow t} d[u'(c_s)/u'(c_t)]} \end{eqnarray}
\begin{eqnarray} \sigma[c(t)] = - \frac{ \lim_{s \rightarrow t} \frac{c_t (dc_s/dc_t) dc_t - c_s dc_t}{c^2_t} }{ \lim_{s \rightarrow t} \frac{ u'(c_t)[du'(c_s)/du'(c_t)] u''(c_t) dc_t - u'(c_s) [u''(c_t)] dc_t}{[u'(c_t)]^2} } \end{eqnarray}
\begin{eqnarray} \sigma[c(t)] = - \frac{ \lim_{s \rightarrow t} - (c_s/c_t) (dc_t/c_t) }{ \lim_{s \rightarrow t} - [u'(c_s)/u'(c_t)] [u''(c_t)/u'(c_t)] dc_t }  \end{eqnarray}
\begin{eqnarray} \sigma[c(t)] = - \frac{ \lim_{s \rightarrow t} (dc_t/c_t) }{ \lim_{s \rightarrow t} [u''(c_t)/u'(c_t)] dc_t } \end{eqnarray}
\begin{eqnarray} \sigma[c(t)] = - \frac{ [1/c(t)] }{ u''[c(t)]/u'[c(t)] } \end{eqnarray}
\begin{eqnarray} \sigma[c(t)] = - \frac{ u'[c(t)] }{ c(t) u''[c(t)] } \end{eqnarray}

Since $\epsilon \equiv \{u''[c(t)] c(t) \}/u'[c(t)]$ is the
elasticity of the marginal utility, then:
\begin{eqnarray} \sigma[c(t)] = - \frac{ 1 }{ \epsilon } \end{eqnarray}

\textbf{Markovian Nash equilibrium}

The Markovian Nash optimal consumption rule at $t$ follows directly
from first-order condition (\ref{foc1}).

The Markovian Nash optimal appropriation rule for $a_i$ can be
derived directly from equations (\ref{d_pi}) and (\ref{d1_pi0}). And
then by simplifying first-order condition (\ref{foc2}):
\begin{eqnarray} - \frac{(1 + \theta a_i)^2}{\theta} =\delta x_j \ \Big(\frac{\gamma \lambda_{ii} - \lambda_{ij}}{\lambda_{ii}}\Big) \end{eqnarray}
\begin{eqnarray} (1 + \theta a_i)^2 = - \theta \delta x_j \ \Big(\frac{\gamma \lambda_{ii} - \lambda_{ij}}{\lambda_{ii}}\Big) \end{eqnarray}
\begin{eqnarray} (1 + \theta a_i)^2 = \theta \delta x_j \ \Big(\frac{\lambda_{ij} - \gamma \lambda_{ii}}{\lambda_{ii}}\Big) \end{eqnarray}
\begin{eqnarray} 1 + \theta a_i = \pm \sqrt{\theta \delta x_j \ \Big(\frac{\lambda_{ij} - \gamma \lambda_{ii}}{\lambda_{ii}}\Big)} \end{eqnarray}
\begin{eqnarray} \theta a_i = \pm \sqrt{\theta \delta x_j \ \Big(\frac{\lambda_{ij} - \gamma \lambda_{ii}}{\lambda_{ii}}\Big)} - 1 \end{eqnarray}
\begin{eqnarray} a_i = \frac{1}{\theta } \Big[\pm \sqrt{\theta \delta x_j \ \Big(\frac{\lambda_{ij} - \gamma \lambda_{ii}}{\lambda_{ii}}\Big)} - 1 \Big]\end{eqnarray}

The nonnegativity condition for $a_i$ requires that:
\begin{eqnarray} \pm \sqrt{\theta \delta x_j \ \Big(\frac{\lambda_{ij} - \gamma \lambda_{ii}}{\lambda_{ii}}\Big)} \geq 1\end{eqnarray}

The plus-minus sign can be dropped, since only the positive square
root case satisfies the condition:
\begin{eqnarray} \sqrt{\theta \delta x_j \ \Big(\frac{\lambda_{ij} - \gamma \lambda_{ii}}{\lambda_{ii}}\Big)} \geq 1\end{eqnarray}
\begin{eqnarray} \theta \delta x_j \ \Big(\frac{\lambda_{ij} - \gamma \lambda_{ii}}{\lambda_{ii}}\Big) \geq 1\end{eqnarray}
\begin{eqnarray} \frac{\lambda_{ij} - \gamma \lambda_{ii}}{\lambda_{ii}} \geq \frac{1}{\theta \delta x_j } \end{eqnarray}

\textbf{Steady state}

The growth rate of the Markovian Nash optimal consumption rule is:
\begin{eqnarray}
\hat{c}^*_i = \sigma(c^*_i) \Big\{ \delta ( 1 - p^*_j) \Big[\frac{
\lambda_{i} - \gamma \lambda_{j} }{\lambda_{i}} \Big] - y'(x_i)
\Big\} \label{grcons_b}
\end{eqnarray}

If $\hat{c}^*_i = 0$, then:
\begin{eqnarray}
\sigma(c^*_i) \Big\{ \delta ( 1 - p^*_j) \Big[\frac{ \lambda_{i} -
\gamma \lambda_{j} }{\lambda_{i}} \Big] - y'(x_i) \Big\} = 0
\end{eqnarray}
\begin{eqnarray}
\delta ( 1 - p^*_j) \Big[\frac{ \lambda_{i} - \gamma \lambda_{j}
}{\lambda_{i}} \Big] - y'(x_i) = 0
\end{eqnarray}
\begin{eqnarray}
\delta ( 1 - p^*_j) \Big[\frac{ \lambda_{i} - \gamma \lambda_{j}
}{\lambda_{i}} \Big] = y'(x_i)
\end{eqnarray}
\begin{eqnarray}
\delta ( 1 - p^*_j) [\lambda_{i} - \gamma \lambda_{j} ] =
\lambda_{i} y'(x_i) \label{eurek}
\end{eqnarray}

\textbf{Wealth inequality}

This shows that a player's stock of wealth and her instantaneous
shadow price are negatively related.  From equation (\ref{eurek}) :
\begin{eqnarray}
y'(x_i)= \delta ( 1 - p^*_j) \Big[\frac{\lambda_{i} - \gamma
\lambda_{j}  }{\lambda_{i}} \Big]
\end{eqnarray}
\begin{eqnarray}
y'(x_i)= \delta ( 1 - p^*_j) \Big[1 - \frac{\gamma \lambda_{j}
}{\lambda_{i}} \Big]
\end{eqnarray}

The partial derivative of the marginal output with respect to the
stock of wealth is:
\begin{eqnarray}
\frac{\partial [y'(x_i)]}{\partial x_i} = - \delta ( 1 - p^*_j)
\gamma \lambda_{j} \Big[\frac{\partial (1/\lambda_{i})}{\partial
x_i} \Big]
\end{eqnarray}
\begin{eqnarray}
y''(x_i) = - \delta ( 1 - p^*_j) \gamma \lambda_{j}
\Big[\frac{\partial (1/\lambda_{i})}{\partial x_i} \Big]
\end{eqnarray}

By assumption, $y''(x_i) > 0$ and $\delta ( 1 - p^*_j) \gamma
\lambda_{j} \geq 0$. Therefore:
\begin{eqnarray}
\frac{\partial (1/\lambda_{i})}{\partial x_i} = - \frac{\partial
\lambda_i/\partial x_i}{\lambda^2_{i}} \leq 0
\end{eqnarray}
\begin{eqnarray}
- \frac{\partial \lambda_i/\partial x_i}{\lambda^2_{i}} \leq 0
\end{eqnarray}

\noindent which implies that $x_i$ and $\lambda_{i}$ are negatively
related.


\newpage

\begin{thebibliography}{10}

\bibitem{adro} Adelman, Irma and Sherman Robinson, ``Income
Distribution and Development,'' \emph{Handbook of Development
Economics}, vol. II, edited by H. Chenery and T.N. Srinivasan, 1989,
Elsevier.

\bibitem{admo} Adelman, Irma and C. T. Morris, \emph{Society,
Politics, and Economic Development: A Quantitative Approach},
Baltimore: John Hopkins Press, 1967.

\bibitem{agbo} Aghion, Phillipe ``A Theory of Trickle-Down Growth and
Development,'' \emph{Review of Economic Studies}, 64, 151-172, 1997.

\bibitem{alro} Alesina, Alberto, and Dani Rodrik, ``Distributive
Politics and Economic Growth,'' \emph{The Quarterly Journal of
Economics}, May 1994.

\bibitem{ar} Arrow, Kenneth J., \emph{Social Choice and Individual Values}, New Haven, CT: Yale
University Press, 1963.

\bibitem{ax} Axelrod, Robert, \emph{The Evolution of Cooperation}, New York: Basic Books,
1984.

\bibitem{babogi} Bardhan, Pranab, Samuel Bowles, and Herbert Gintis,
``Wealth Inequality, Wealth Constraints and Economic Performance,''
Unpublished Paper, 1998 (forthcoming in A. Atkinson and F.
Bourguignon (Eds), \emph{Handbook of Income Distribution},
North-Holland).

\bibitem{ba} Barro, Robert J., ``Inequality, Growth, and
Investment,'' NBER Working Paper No. w7038, March 1999.

\bibitem{basa} Barro, Robert and Xavier Sala-i-Martin,
\emph{Economic Growth}, McGraw-Hill, 1995.

\bibitem{be} B\'{e}nabou, R., ``Inequality and Growth”, in Bernanke, B. and J. Rotemberg (editors),
NBER Macro Annual 1996, MIT Press: Cambridge, MA, pp. 11-76, 1996.

\bibitem{be} Bertola, Giuseppe, ``Macroeconomics of Distribution
and Growth,'' \emph{Handbook of Income Distribution}, vol. I, edited
by A.B. Atkinson and F. Bourguignon, 1999, Elsevier.

\bibitem{br1} Bertkovicz, L.D., ``Necessary Conditions for Optimal Strategies in a Class of
Differential Games and Control Problems,'' \emph{Annals of
Mathematics}, Study No. 52, ed. M. Dresher, L.S. Shapley, and A.W.
Tucker, Princeton, NJ: Princeton University Press, 1964.

\bibitem{br2} Bertkovicz, L.D., ``A Survey of Differential Games,'' in A.V. Balakrishnan and
L.W. Neustadt (editors), \emph{Mathematical Theory of Control}, New
York: Academic Press, 1967.

\bibitem{bilo} Birdsall, N. and J.L. Londono, ``Asset Inequality Matters: An Assessment of the World
Bank's Approach to Poverty Reduction,'' \emph{American Economic
Review}, Vo. 82, No. 2, pp. 32-37, 1997.

\bibitem{blni} Bleaney, Michael and Akira Nishiyama, ``Explaining
Growth: A Contest Between Models,'' \emph{Journal of Economic
Growth}, 7, 259-281, 2002, Norwell, Mass.: Kluwer Academic
Publishers.

\bibitem{bogi} Bowles, Samuel and Herbert Gintis, ``The Evolution of Strong Reciprocity,''
Unpublished Paper, University of Massachusetts Amherst, 2000.

\bibitem{chta} Chou, Chien Fu and Gabriel Talmain, ``Redistribution and Growth: Pareto
Improvements,'' \emph{Journal of Economic Growth}, 1: 505-523,
December 1996.

\bibitem{clwa} Clemhout, S. and H.Y. Wan, Jr., ``Differential Games: Economic Applications,''
in Aumann, R.J. and S. Hart (editors), \emph{Handbook of Game Theory
with Economic Applications}, Vol. II, Amsterdam: Elsevier, 1994.

\bibitem{co} Coase, Ronald, ``The Problem of Social Cost,'' \emph{Journal of Law and Economics}, 1960,
republished in Ronald H. Coase, \emph{The Firm, the Market and the
Law}, The University of Chicago Press, 1988.

\bibitem{cl} Clarke, George R. G., ``More Evidence on Income
Distribution and Growth,'' University of Rochester, Unpublished
Paper, June 1993.

\bibitem{coli} Conley, Timothy G. and Ethan Ligon ``Economic
Distance and Cross-Country Spillovers,'' \emph{Journal of Economic
Growth}, 7, 157-187, 2002, Norwell, Mass.: Kluwer Academic
Publishers.

\bibitem{de} Decornez, S. S. ``An Empirical Analysis of the
American Middle Class (1968-1992),'' Ph.D. Dissertation, Vanderbilt
University.

\bibitem{desq} Deininger, K. and L. Squire, ``A New Data Set Measuring Income Inequality,''
\emph{World Bank Economic Review}, Vol. 10, pp. 565-591, 1996.

\bibitem{desq1} Deininger, K. and L. Squire, ``New Ways of Looking at Old Issues,'' \emph{Journal of
Development Economics}, Vol. 57, pp. 259-87, 1998.

\bibitem{dojo} Dockner, Engelbert J., Steffen J\o rgensen, Ngo Van Long, and Gerhard Sorger,
\emph{Differential Games in Economics and Management Science},
Cambridge: Cambridge University Press, 2000.

\bibitem{ea} Easterly, William, ``The Middle Class Consensus and
Economic Development,'' \emph{Journal of Economic Growth}, 6,
317-335, 2001, Norwell, Mass.: Kluwer Academic Publishers.

\bibitem{ea2} Easterly, William, ``Inequality does Cause Underdevelopment: New Evidence from
Commodity Endowments, Middle Class Share, and Other Determinants of
Per Capita Income,'' Preliminary Paper, Washington, DC: Development
Research Group, World Bank, 2001.

\bibitem{eale} Easterly, William and R. Levine, ``Africa's Growth
Tragedy: Policies and Ethnic Divisions,'' \emph{The Quarterly
Journal of Economics}, 112 (4), 1203-1250.

\bibitem{enso} Engerman, Stanley L. and Kenneth L. Sokoloff, \emph{Factor Endowments,
Inequality, and Paths of Development Among New World Economies},
NBER, Working Paper 9259, October 2002.

\bibitem{fe} Feldstein, Martin, ``Social Insurance,'' in \emph{Income
Redistribution}, Colin D. Campbell (editor), Washington, DC:
American Enterprise Institute for Public Policy Research, 1976.

\bibitem{fer} Ferreira, Francisco H.G., ``Inequality and Economic Performance: A Brief Overview
to Theories of Growth and Distribution,'' Text for World Bank’s Web
Site on Inequality, Poverty, and Socio-economic Performance,
http://www.worldbank.org/poverty/inequal/index.htm, June 1999.

\bibitem{fisi} Fishman, Arthur and Avi Simhon, ``The Division of
Labor, Inequality, and Growth,'' \emph{Journal of Economic Growth},
7, 117-136, 2002, Norwell, Mass.: Kluwer Academic Publishers.

\bibitem{fo} Forbes, Kristin J., “A Reassessment of the Relationship Between Inequality and Growth”,
\emph{American Economic Review}, Vol. 90, No. 4, pp. 869-887,
September 2000.

\bibitem{fr} Friedman, A., \emph{Differential Games}, New York: Wesley \& Son, 1971.

\bibitem{fr3} Friedman, Milton, \emph{Capitalism and Freedom}, The University of Chicago Press, 1962.

\bibitem{gaze} Galor, Oded and Joseph Zeira, ``Income Distribution and Macroeconomics,'' \emph{Review of
Economic Studies}, 60, pp. 35-52, 1993.

\bibitem{grki} Grossman, Herschel I. and Minseong Kim, ``Swords
or Plowshares? A Theory of the Security of Claims to Property,''
\emph{Journal of Political Economy}, 1995, Vol. 103, No. 6.

\bibitem{ha} Haavelmo, Trygve, \emph{A Study in the Theory of Economic
Evolution}, Amsterdam: North-Holland, 1954.

\bibitem{har} Harrod, Roy F., \emph{Towards a dynamic economics},
London: Macmillan, 1948.

\bibitem{ho} Hoel, Michael, ``Distribution and Growth as a Differential Game between Workers and Capitalists,'' \emph{International Economic Review},
Vol. 19, No. 2, June 1978.

\bibitem{in} Intriligator, Michael D., \emph{Mathematical Optimization and
Economic Theory}, New York: Prentice-Hall, 1971.

\bibitem{is} Isaacs, Rufus, \emph{Differential Games: A Mathematical Theory with Applications to Warfare and
Pursuit, Control and Optimization}, New York: Wesley \& Sons.

\bibitem{is} Itaya, Yuji, ``Dynamic Optimization and Differential Games with Applications to Economics,'' Manuscript,
Gifu, Japan, 2000.

\bibitem{kal} Kalemli-Ozcan, Sebnem ``Does the Mortality Decline
Promote Economic Growth?'' \emph{Journal of Economic Growth}, 7,
411-439, 2002, Norwell, Mass.: Kluwer Academic Publishers.

\bibitem{kan} Kanbur, Ravi, ``Income Distribution and
Development,'' \emph{Handbook of Income Distribution}, vol. I,
edited by A.B. Atkinson and F. Bourguignon, 2000, Elsevier.

\bibitem{krsu} Krasovskii, N.N. and A.I. Subbotin, \emph{Game-Theoretical Control Problems},
New York: Springer-Verlag, 1990.

\bibitem{kr} Kremer, Michael ``Income Distribution Dynamics with
Endogenous Fertility'' \emph{Journal of Economic Growth}, 7,
227-258, 2002, Norwell, Mass.: Kluwer Academic Publishers.

\bibitem{kri} Kristol, Irving, ``Thoughts on Equality and Egalitarianism,'' in \emph{Income
Redistribution}, Colin D. Campbell (editor), Washington, DC:
American Enterprise Institute for Public Policy Research, 1976.

\bibitem{lac} Lancaster, Kelvin, ``The Dynamic Inefficiency of Capitalism,'' \emph{Journal of
Political Economy},  Vol. 81, No. 5, September-October 1973.

\bibitem{la} Landes, David, \emph{The Wealth and Poverty of
Nations}, New York, NY: Norton, 1998.

\bibitem{lay} Layard, Richard, ``Happiness: Has Social Science a
Clue?'' Lionel Robbins Memorial Lectures 2002/3, London School of
Economics, March 2003.

\bibitem{le} Leitmann, G., \emph{Cooperative and Non-cooperative Many Players Differential Games},
New York: Springer-Verlag, 1974.

\bibitem{lizo} Li, Hongyi and Heng-fu Zou, ``Income Inequality is not Harmful for Growth: Theory and
Evidence,'' \emph{Review of Development Economics}, Vol. 2, No. 3,
October 1998.

\bibitem{lu} Lucas, Robert E., Jr., ``On the Mechanics of Economic Development,'' \emph{Journal of
Monetary Economics}, 22, July 1988.

\bibitem{marowe} Mankiw, N. Gregory, David Romer, and D. Weil,
``A Contribution to the Empirics of Economic Growth,''
\emph{Quarterly Journal of Economics}, 101, 407-437, 1992.

\bibitem{mar} Marx, Karl, \emph{Capital: A Critique of Political Economy}, volumes 1, New York:
Vintage Books, 1977.

\bibitem{max} Marx, Karl, \emph{Grundrisse: Foundations of the Critique of Political Economy
(Rough Draft)}, Harmondsworth, England: Penguin, 1973.

\bibitem{ma} Marx, Karl, \emph{Critique of the Gotha Program}, in K. Marx and F. Engels,
\emph{Selected Works}, volume II, Moscow: International Publishers,
1938.

\bibitem{maen} Marx, Karl and Friederich Engels, \emph{Manifesto of
the Communist Party}, 1848, in \emph{Collected Works}, volume 6,
1976, New York: International Publishers.

\bibitem{mc} McDermott, John ``Development Dynamics: Economic
Integration and the Demographic Transition,'' \emph{Journal of
Economic Growth}, 7, 371-409, 2002, Norwell, Mass.: Kluwer Academic
Publishers.

\bibitem{me} Mendez, Rodrigue ``Creative Destruction and the Rise of Inequality,''
\emph{Journal of Economic Growth}, 7, 259-281, 2002, Norwell, Mass.:
Kluwer Academic Publishers.

\bibitem{ok} Okun, Arthur M. \emph{Equality and Efficiency: The Big Tradeoff},
Washington: The Brookings Institution, 1975.

\bibitem{ok1} Okun, Arthur M. ``Further Thoughts on Equality and Efficiency,'' in \emph{Income
Redistribution}, Colin D. Campbell (editor), Washington, DC:
American Enterprise Institute for Public Policy Research, 1976.

\bibitem{pa} Panizza, Ugo ``Income Inequality and Economic
Growth: Evidence from American Data,'' \emph{Journal of Economic
Growth}, 7, 25-41, 2002, Norwell, Mass.: Kluwer Academic Publishers.

\bibitem{pe} Perotti, R. ``Political Equilibrium, Income Distribution, and Growth,''
\emph{Review of Economic Studies}, 60, 1993, pp. 755-776.

\bibitem{pe1} Perotti, R. ``Growth, Income Distribution, and Democracy: What the Data Say''
\emph{Journal of Economic Growth}, Vol. 1, No. 2, pp. 149-187, June
1996.

\bibitem{peta1} Persson T. and G. Tabellini, ``Is Inequality harmful for Growth? Theory and
Evidence,'' \emph{American Economic Review}, Vol. 84, No. 3, pp.
600-621, 1994.

\bibitem{pet} Petit, M.L., \emph{Control Theory and Dynamic Games in Economic Policy Analysis},
Cambridge: Cambridge University Press, 1990.

\bibitem{pi} Pigou, A.C., \emph{The economics of Welfare}, Fourth Edition,
London: Macmillan, 1932.

\bibitem{pi} Pontryagin, L.S., et al. \emph{The Mathematical Theory of Optimal Processes},
Translation: K. N. Trirogoff, Edition: L. W. Neustadt, New York:
Wiley-Interscience, 1962.

\bibitem{qu} Quah, Danny, ``One Third of the World's Growth and
Inequality,'' Unpublished Paper, London School of Economics, April
2002.

\bibitem{ra} Ramsey, Frank P., ``A Mathematical Theory of
Saving,'' \emph{Economic Journal}, 38, 1928.

\bibitem{sa} Sala-i-Martin, Xavier ``Transfers, Social Safety
Nets, and Economic Growth'' \emph{IMF Staff Papers}, Vol. 44,
81-102, March 1997, International Monetary Fund.

\bibitem{se} Sen, Amartya \emph{On Economic Inequality}, Expanded Edition, 1997,
Oxford: Clarendon Press.

\bibitem{se2} Sen, Amartya \emph{Inequality Reexamined}, Cambridge, Mass.: Harvard University Press, 1995.

\bibitem{sh} Shiller, Robert, \emph{Macro Markets}, TIAA-CREF Paul Samuelson Award, New York: Oxford
University Press, 1996.

\bibitem{soen} Sokoloff, Kenneth L. and Stanley L. Engerman, Institutions, Factor Endowments, and
Paths of Development in the New World, \emph{Journal of Economic
Perspectives}, Vol. 14, No. 3, 217-32, 2000.

\bibitem{stlu} Stokey, Nancy L. and Rober E. Lucas, Jr. (with Edward C. Prescott), \emph{Recursive
Methods in Economic Dynamics}, Cambridge, Mass.: Harvard University
Press, 1989.

\bibitem{sti} Stiglitz, Joseph, \emph{The Roaring Nineties}, New York: W.W. Norton \& Co., 2003.

\bibitem{szhi} Sz\'ekely, Miguel and Marianne Hilgert, ``What's Behind the
Inequality We Measure?'' Inter-American Development Bank, Research
Department Working Paper 409, 1999.

\bibitem{taar} Taylor, Lance and Persio Arida, ``Long-Run Income
Distribution and Growth,'' \emph{Handbook of Development Economics},
vol. I, edited by H. Chenery and T.N. Srinivasan, 1988, Elsevier.

\bibitem{to} Tobin, James, ``Considerations Regarding Taxation and Inequality,'' in \emph{Income
Redistribution}, Colin D. Campbell (editor), Washington, DC:
American Enterprise Institute for Public Policy Research, 1976.

\end{thebibliography}
\end{document}